\documentclass[aps,prl,superscriptaddress,twocolumn]{revtex4-1}
\usepackage{graphicx}
\usepackage{enumitem}

\usepackage{xr-hyper}  
\usepackage{ulem}  
\usepackage{soul}  
\usepackage{comment}   
\usepackage{afterpage}  

\usepackage{hyperref}
\usepackage{amsmath}
\usepackage{cleveref}
\usepackage{amssymb}
\usepackage{xcolor}  
\usepackage{xfp}

\begin{document}


\title{Hysteretic effects and magnetotransport of electrically switched CuMnAs}




\author{Jan~Zub{\'a}{\v c}}
\email{zubac@fzu.cz}
\affiliation{Institute of Physics, Czech Academy of Sciences,
Cukrovarnick{\'a} 10, 162 00, Prague 6, Czech Republic}
\affiliation{Charles University, Faculty of Mathematics and Physics, Ke Karlovu 3, 121 16 Prague 2, Czech Republic}

\author{Zden{\v e}k~Ka{\v s}par}
\affiliation{Institute of Physics, Czech Academy of Sciences,
Cukrovarnick{\'a} 10, 162 00, Prague 6, Czech Republic}
\affiliation{Charles University, Faculty of Mathematics and Physics, Ke Karlovu 3, 121 16 Prague 2, Czech Republic}

\author{Filip~Krizek}
\affiliation{Institute of Physics, Czech Academy of Sciences,
Cukrovarnick{\'a} 10, 162 00, Prague 6, Czech Republic}

\author{Tobias~F{\"o}rster}
\affiliation{Hochfeld-Magnetlabor Dresden (HLD-EMFL) and W\"urzburg-Dresden Cluster of Excellence ct.qmat, Helmholtz-Zentrum
Dresden-Rossendorf, 01328 Dresden, Germany}

\author{Richard~P.~Campion}
\affiliation{School of Physics and Astronomy, University of Nottingham,Nottingham NG7 2RD, United Kingdom}

\author{V{\'i}t~Nov{\'a}k}
\affiliation{Institute of Physics, Czech Academy of Sciences,
Cukrovarnick{\'a} 10, 162 00, Prague 6, Czech Republic}

\author{Tom{\'a}{\v s}~Jungwirth}
\affiliation{Institute of Physics, Czech Academy of Sciences,
Cukrovarnick{\'a} 10, 162 00, Prague 6, Czech Republic}
\affiliation{School of Physics and Astronomy, University of Nottingham,Nottingham NG7 2RD, United Kingdom}

\author{Kamil~Olejn{\'i}k}
\email{olejnik@fzu.cz}
\affiliation{Institute of Physics, Czech Academy of Sciences,
Cukrovarnick{\'a} 10, 162 00, Prague 6, Czech Republic}

\date{\today}

\begin{abstract}
Antiferromagnetic spintronics allows us to explore storing and processing information in magnetic crystals with vanishing magnetization. In this manuscript, we investigate magnetoresistance effects in antiferromagnetic CuMnAs upon switching into high-resistive states using electrical pulses.
By employing magnetic field sweeps up to 14~T and magnetic field pulses up to $\sim 60$ ~T, we reveal hysteretic phenomena and changes in the magnetoresistance, as well as the resilience of the switching signal in CuMnAs to the high magnetic field. 
These properties of the switched state are discussed in the context of recent studies of antiferromagnetic textures in CuMnAs.

\end{abstract}

\maketitle

\section{I.~Introduction}
\label{Section I.}

Antiferromagnetic materials gained attention in contemporary spintronics due to their unique features: The fast terahertz magnetization dynamics of antiferromagnets overcomes the GHz limit of their ferromagnetic counterparts used in present-day microelectronics  (e.g. in magnetic random-access memories - MRAMs). Zero net magnetization due to the alternating magnetic moments on neighbouring atoms and the resulting absence of stray fields are advantageous for memory applications requiring high integration density with no cross-talk among adjacent devices \cite{Jungwirth2016Antiferromagnetic, Baltz2018Antiferromagnetic}.
On the other hand, antiferromagnets have low sensitivity to external magnetic fields. This limits the implementation of schemes used for the manipulation of magnetic moments in ferromagnets and has, until recently, hindered the research into practical applications of antiferromagnetic materials.

One of the possible methods of controlling the antiferromagnetic order is by N{\'e}el  spin-orbit torques, which were predicted in materials with specific symmetries.
For example, an efficient N{\'e}el  vector reorientation can be generated by a field-like spin-orbit torque when
the antiferromagnetically coupled magnetic atoms occupy inversion-partner lattice sites \cite{Zelezny2014Relativistic}.
A tetragonal collinear antiferromagnet CuMnAs with the N{\'e}el  temperature $T_{\mathrm{N}} = 480$~K \cite{Wadley2013Tetragonal, Wadley2015Antiferromagnetic} is an example of a material where this switching mechanism was experimentally demonstrated \cite{Wadley2016Electrical, Wadley2018Current}. In this material,
electrical control of the antiferromagnetic moments was also detected by the anisotropic magnetoresistance (AMR) \cite{Wadley2016Electrical}. 
However, the AMR-related signals were rather small, with relative resistance changes limited to tenths of percent \cite{Wadley2016Electrical, Wadley2018Current, Volny2020Electrical, Wang2020Spin}.
Apart from CuMnAs, the electrical control of magnetic moments was also demonstrated in Mn$_2$Au \cite{Bodnar2018Writing, Bodnar2019Imaging, Meinert2018Electrical} and other antiferromagnetic  materials \cite{Nair2020Electrical,Cheng2020Electrical,Baldrati2019Mechanism, Dunz2020Spin}. 

Recently, we reported switching of CuMnAs into high-resistive states using optical or unipolar electrical pulses with resistive switching signals reaching 20$\%$ at room temperature
and approaching 100$\%$ at low temperatures \cite{Kaspar2020Quenching}. 
After the writing pulse, the resistive signals relax following a stretched-exponential time-dependence with a characteristic timescale showing a simple exponential dependence on temperature.
Unlike the spin-orbit torque reorientation of the N{\'e}el  vector controlled by the angle or polarity of the switching current, the unipolar switching mechanism is independent of the writing current direction or polarization of the switching laser pulses.
Complementary imaging measurements, including 
X-ray magnetic linear dichroism photoemission electron microscopy 
(XMLD-PEEM) and
magneto-Seebeck microscopy \cite{Kaspar2020Quenching, Janda2020Magneto}, 
have shown that the observed large switching signals were accompanied by  nano-fragmentation of magnetic domains. 
This unipolar mechanism, which is principally distinct from the spin-orbit torque reorientation of the N{\'e}el vector,
was called a quench switching.
In addition, recent imaging by differential-phase-contrast scanning transmission electron microscopy (DPC-STEM) \cite{krizek2020atomically}
has revealed that complex magnetic textures in CuMnAs can even include atomically sharp  180$^{\circ}$ domain walls. 

Effects of the applied magnetic field on CuMnAs films and devices have been previously studied \cite{Wang2020Spin} by means of magnetotransport measurements, XMLD spectroscopy and XMLD-PEEM. The study demonstrated a spin-flop transition and spin reorientation at the magnetic field of $\sim 2$~T in CuMnAs thin films with uniaxial and biaxial magnetic anisotropy, respectively.
Measurements in the spin-flop (reorientation) fields confirmed that AMR associated with the reorientation of the N{\'e}el vector is on the $\sim 0.1\%$ scale in CuMnAs.

In this manuscript, we investigate the quenched high-resistive states of CuMnAs at strong magnetic fields. Magnetoresistance (MR), i.e. the dependence of the material's resistance on the applied magnetic field, is a straightforward method to examine both magnetic and transport characteristics of metallic thin films.
This phenomenon can be of multiple origins but
the following should be especially taken into consideration
for antiferromagnetic CuMnAs:
(i)~The above-mentioned anisotropic magnetoresistance (AMR),
(ii)~ordinary metallic MR caused by the Lorentz force,
(iii)~spin-disorder contribution
to the resistance, and (iv)~domain wall MR which can substantially affect the device resistance in particular when containing the sharp domain walls of high densities \cite{Kent2001Domain}.

Our paper is organized as follows: 
Section II describes employed experimental methods.
In Section III, we present the main results: We show the qualitative differences between the MRs of the quenched high-resistive state and
the relaxed state (\autoref{fig2}), and the correspondence between the size of the switching signal and the magnitude of the changes detected in the MR (\autoref{fig3new}). We also address the anisotropy of the observed effects (\autoref{fig4}) and the behaviour at extreme magnetic fields (\autoref{fig5}). In Section IV, we discuss our MR results in the context of the earlier magnetic-texture microscopy studies. Finally, Section V summarizes our results.

\section{II.~Experimental methods}
 Tetragonal CuMnAs films of thickness 50~nm and 20~$\Omega$ sheet resistance were grown by molecular beam epitaxy on GaP(001) substrates and capped with 3~nm of Al to prevent oxidation. Characterization by atomic force and scanning electron microscopy, SQUID magnetometry, x-ray diffraction and transport measurements showed high crystal quality and optimal electrical switching performance of the prepared samples. 
In particular, SQUID magnetometry data, which are typically 
dominated by a large negative diamagnetic signal of the GaP substrate,
do not show any substantial magnetic moment after subtraction of this linear diamagnetic contribution. Further details about sample growth and characterization were recently described in Ref.~\cite{Krizek2020Molecular}. 

Hall bars (width 10$\mu$m, length 50$\mu$m) used in experiments were patterned by electron beam lithography and wet etching (see inset of \autoref{fig1}~c for a scanning electron microscopy image of the fabricated device).

Magnetoresistance measurements up to 14 T were performed using the electrical transport option (ETO) of the Physical Property Measurement System (PPMS). The system utilizes the digital lock-in technique and is equipped with a rotator for in-plane angular scans. Selected samples were later exposed to the magnetic field pulses with the magnitude of $\sim 60$~T and the total duration of $\sim 150$~ms at the Dresden High Magnetic Field Laboratory. 
Resistance evolution during the magnetic field pulse was recorded by a
high-speed digitizer, the obtained data were processed using a lock-in technique.

\section{III.~Results}

\subsection*{A.~Electrical switching and the magnetotransport measurement protocol}

\begin{figure}[h]
 \centering
 \includegraphics[width=\textwidth]{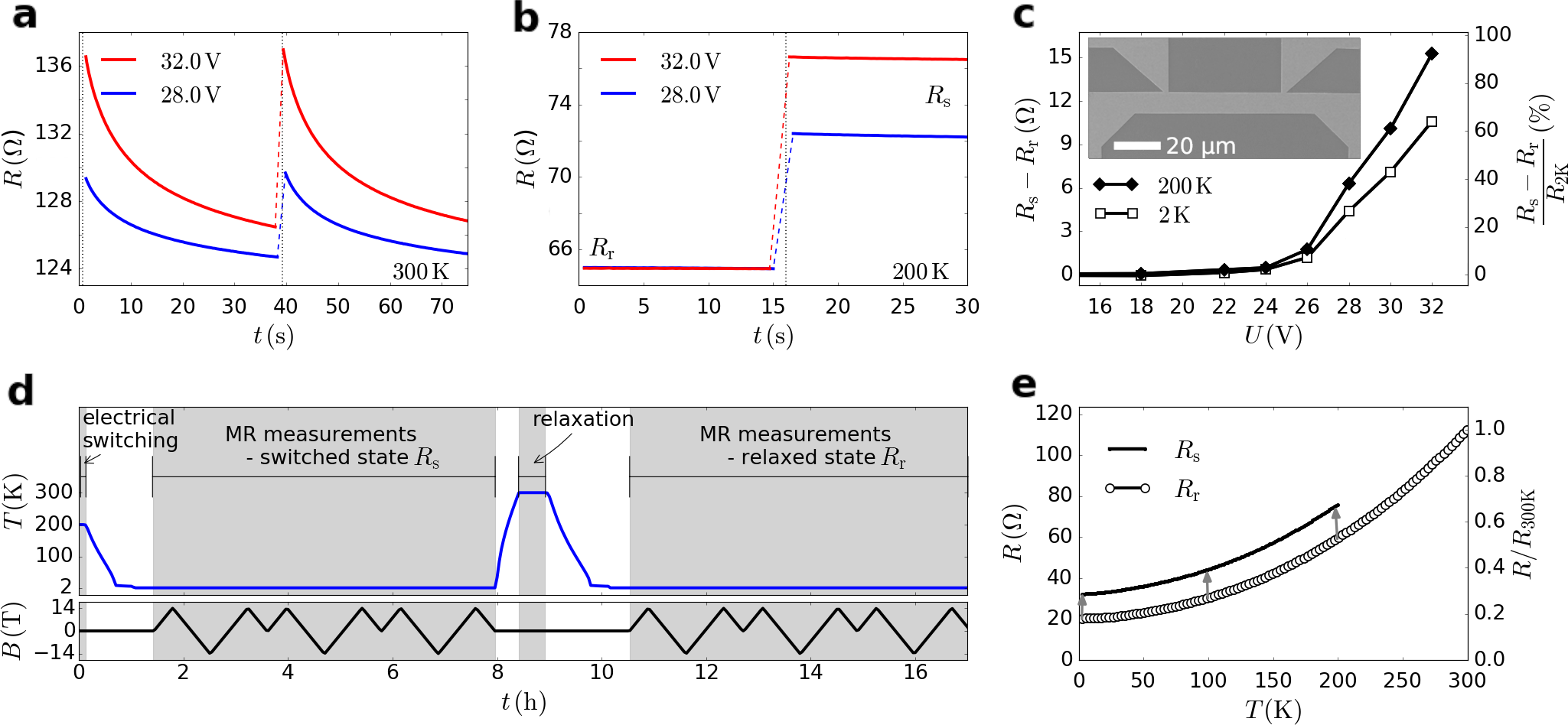}
 \caption{ \textbf{(a)}~Electrical switching of a typical Hall bar device at 300~K showing relaxation of a signal. Single unipolar 100$\mu$s writing pulses of two different voltage amplitudes and longitudinal resistance readout were employed. The dotted lines mark the positions of the individual pulses.	
           \textbf{(b)}~Non-relaxing resistance signal after the electrical switching from a relaxed ($R_{\mathrm{r}}$) to the switched ($R_{\mathrm{s}}$)
           state at 200~K.
           \textbf{(c)}~Magnitude of the signal induced by electrical switching at 200~K evaluated at 200~K during switching (Fig.~1b) and at 2~K as the difference between the switched resistance and the resistance after relaxation (see Fig.~1d). The highest pulse voltage of 32~V corresponds to the current density of $\sim1.2\times 10^7 \,\mathrm{A/cm}^2$.
           The inset shows a SEM micrograph of a typical Hall bar device.
		   \textbf{(d)}~Outline of the main experiment: After electrical switching at 200~K the sample is rapidly cooled down to the temperature of 2~K at which MR field-sweep measurements in the quenched switched state are performed. Then the sample is warmed up and kept at the temperature of 300~K for 30~min to recover its equilibrium relaxed state. Afterwards, MR in the relaxed state at 2~K is measured in the same manner and with the same field sweep rate of 0.66~T/min as in the switched state.
           \textbf{(e)}~Temperature dependence of resistance of a typical Hall bar device measured in the relaxed (black symbols) and switched state (grey symbols). Arrows are used to denote a shift of resistance by $\Delta R = R_{\mathrm{s}} - R_{\mathrm{r}}$ due to the electrical switching by 30~V pulses at 200~K. Both curves were obtained during heating for the current applied along the $[\overline{1}00]$ CuMnAs direction.}
           
 \label{fig1}
\end{figure}

In \autoref{fig1}a,~b,~c  we demonstrate the switching of our CuMnAs Hall bars by unipolar electrical pulses of distinct voltage amplitudes and at different base temperatures. Whereas at 300~K, the resistance after the electrical pulse noticeably decays towards the initial low value (\autoref{fig1}a), a high-resistance state is preserved and negligible relaxation is observed after the pulsing at 200~K (\autoref{fig1}b). Besides this, the resistive signal scales up with the applied pulse voltage as illustrated in \autoref{fig1}c for the switching at 200~K.
In Ref.~\cite{Kaspar2020Quenching}, it was shown that the signal induced by the electrical switching consists of two main components 
with different relaxation times $\tau_1$ and $\tau_2$. Both components follow the stretched exponential function $\sim \exp( - (t/\tau_{1(2)})^{0.6} )$ and their relaxation times depend exponentially on temperature as $\tau_{1(2)} = \tau_0 \exp(E_{1(2)}/ k_{ \mathrm{B} } T)$,
consistently with relaxation behaviour recognized in many complex systems such as glassy materials \cite{dotsenko1991spin, lookman2018frustrated}.
Using the energy parameters $E_1/k_{\mathrm{B}} = 9240$~K, $E_2/k_{\mathrm{B}} = 7830$~K and $1/\tau_0$ in the THz range from Ref.~\cite{Kaspar2020Quenching},
we get $\tau_1 \sim$~10~s for the main switching component and $\tau_2 \sim$~10~ms for the minor fast-relaxing component at 300~K, and  $\tau_1 \sim$~10$^7$~s and $\tau_2 \sim$~10$^4$~s at 200~K for the two components, respectively. 
These expressions and parameters explain the observed non-decaying behaviour
after cooling and allow us to realize MR measurements in the electrically switched state with  virtually infinite relaxation times at low temperatures ($\sim 2$~K).

The design of the main MR experiment is schematically presented in \autoref{fig1}d. At first, the electrical switching is performed at 200~K. Immediately
after the switching, the sample is rapidly cooled down to 2~K maintaining the high resistance of the switched state.
After temperature stabilization, MR curves for the magnetic field applied along selected crystallographic directions are collected. To recover the low-resistance relaxed state, the sample is heated up and annealed at 300~K for 30~minutes ($\gg \tau_{1(2)}$). Subsequently, the sample is again cooled down to 2~K and MR measurements are repeated in the same manner, but this time in the relaxed state.

\autoref{fig1}e shows the temperature dependence of the resistance measured on a typical Hall bar device in the relaxed ($R_{\mathrm{r}}$) and in the switched ($R_{\mathrm{s}}$) state. The relaxed-state resistance exhibits smoothly varying metallic behaviour with a residual resistance ratio $\approx 6$, obtained from the 300 K and 2 K measurements, indicating high crystal quality of the samples. The switched state is offset by $\delta R = R_{\mathrm{s}} - R_{\mathrm{r}}$. This difference does not change substantially in the displayed temperature range from 2 to 200~K.

\subsection*{B.~Field-sweep magnetotransport up to 14T}
\begin{figure}[h]
 \centering
 \includegraphics[width=0.88\textwidth]{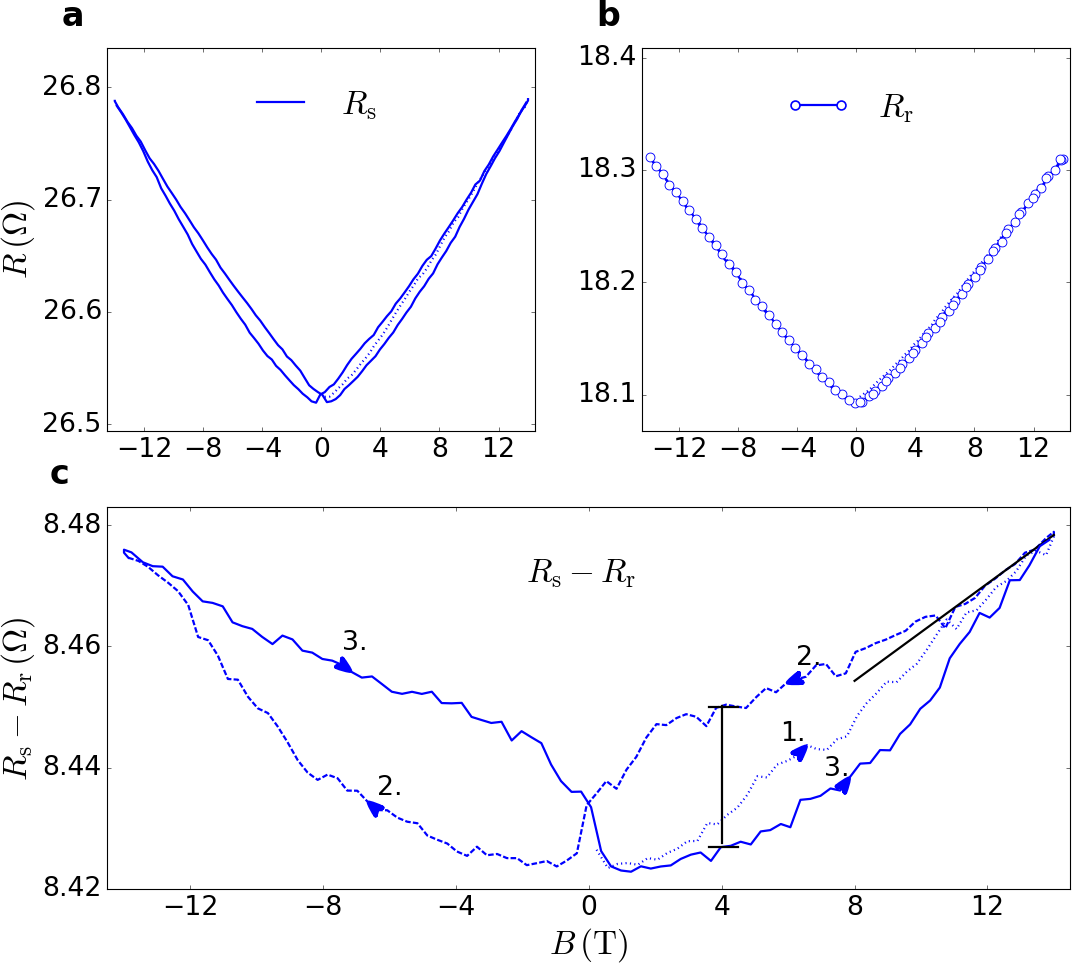}
  \caption{Magnetoresistance of CuMnAs at 2~K \textbf{(a)} in the high-resistance switched  state ($R_{\mathrm{s}}$) for  the switching signal $\sim 8.5\, \Omega$  and \textbf{(b)} in the low-resistance relaxed state  ($R_{\mathrm{r}}$). The data were collected for the $j \parallel B \parallel [\overline{1}00]$ configuration. Dotted lines represent the first sweep up after cooling the sample. \textbf{(c)}~Difference $R_{\mathrm{s}} - R_{\mathrm{r}}$ between the switched-state and the relaxed state magnetoresistance showing main effects induced by the quench switching: significant hysteretic behaviour and the change of the high-field slope. Black lines mark hysteresis magnitude at 4~T and the slope from the linear fit of the sweep down in the range 10-14~T. Arrows and numbers indicate the development of the hysteretic magnetization process. Further discussion in the text.}
 \label{fig2}
\end{figure}

\autoref{fig2} shows main changes in the MR produced by the quench switching.
Whereas the relaxed state MR (\autoref{fig2}b) displays 
essentially identical MRs for both up and down field sweeps, the MR in the high-resistance switched state (\autoref{fig2}a) exhibits pronounced 
hysteretic effects. Moreover, a distinct response of the switched and relaxed state to the magnetic field is reflected in the change of the high field slope of the MR curves.
To highlight the differences between the two states, we subtracted the relaxed-state MR from the switched-state MR, and the result is presented in \autoref{fig2}c.
The evolution of the hysteretic magnetization process and its main features are indicated by arrows and
numbers:
First, the resistance follows the virgin curve during the magnetic field increase (1.).  Second, we record a higher resistance trace from 14~T to 0~T and a subsequent drop in the resistance to a lower value after crossing the zero-field point and continuing to $-14$~T (2.).
The MR curve~(3.) is then recorded when the field is swept from -14~T to 14~T
to complete a butterfly-shaped pattern. Subsequent magnetic field sweeps in the range from -14~T to 14~T can be described by repeating curves of the type (2.) and (3.) with the crossing at the same remanent resistance in the zero field. No considerable erasing of the overall switching signal by the magnetic field can be identified.
The MR traces show reproducible hysteretic behaviour with a maximum width of the hysteresis at approximately 4~T. Additional measurements of minor hysteresis loops in the switched state of CuMnAs are presented in Supplementary Fig.~S1.

\begin{figure}[h]
 \centering
 \includegraphics[width=\textwidth]{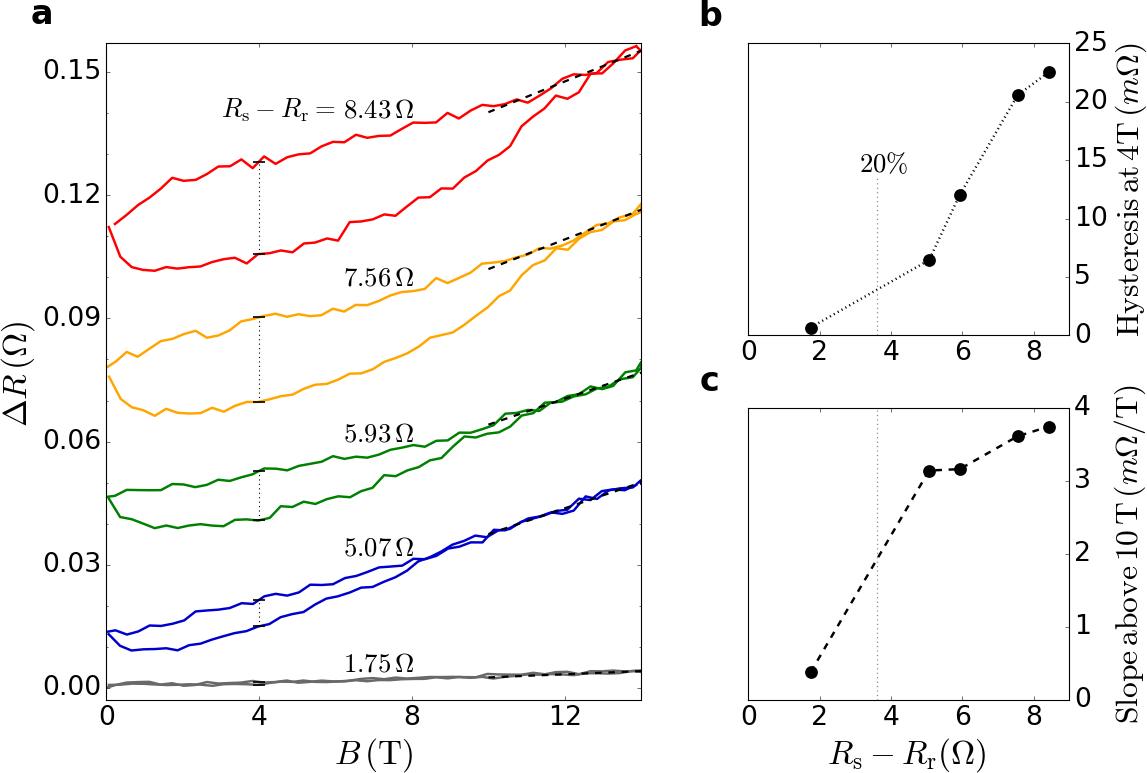}
 \caption{ \textbf{(a)} Difference MR curves between the switched and the relaxed state 
 for various switching signal magnitudes. The data were taken for the $j \parallel B \parallel [\overline{1}00]$ configuration; the virgin MR curve is not shown. Individual dependencies were further shifted by an arbitrary value for lucidity. Development of the hysteresis size~\textbf{(b)} and the high-field slope~\textbf{(c)}  with the switching signal size for the data in \textbf{(a)}. The switching signal $(R_{\mathrm{s}} - R_{\mathrm{r}})/R_{\mathrm{2K}}  = 20 \%$ is indicated by a grey vertical line. 
 }
 		
 \label{fig3new}
\end{figure}

To demonstrate the close connection between the electrical switching and the changes observed in the MR, we repeated the magnetotransport measurements for several magnitudes of the switching signal. The obtained difference MR data are presented in \autoref{fig3new}a. For small switching signals ($\lesssim 20\%$), the switched-state MR roughly follows its relaxed-state counterpart, giving an almost constant difference MR curve. For larger switching signals, we observe the sizeable hysteresis accompanied by the change in the high-field slope (\autoref{fig3new}b,~c). These two effects are the dominant features emerging in the MR after the switching.
Qualitatively similar but quantitatively varying effects were also observed for different mutual current and field configurations (see Supplementary Figs.~S2 and S3 for details).

In \autoref{fig4}, we explore the anisotropy of the switching-induced changes by measuring the longitudinal resistance during magnetic field rotations. The experimental procedure used here was analogous to that of field-sweep measurements.  
In tetragonal systems with competing uniaxial and biaxial anisotropies, the longitudinal AMR can be described by the following expression \cite{Rushforth2007Anisotropic}:
\begin{equation}
  \begin{aligned}
 \Delta R/\overline{R} & = C_{I} \cos( 2\psi ) + C_{U}\cos( 2\phi) \\  
                       & + C_{C}\cos( 4\phi ) + C_{IC}\cos( 4\phi-2\psi ),
  \end{aligned}
    \label{eqAMRxx}
\end{equation}
where $\psi$ is the angle between the N{\' e}el vector and the probing current $j$, $\phi$ is the angle between the N{\' e}el vector and the $[100]$ crystalline direction of CuMnAs, and the individual
terms represent the non-crystalline AMR contribution ($C_{I}$), the uniaxial  ($C_{U}$) and cubic ($C_{C}$) crystalline contribution and the mixed crystalline and non-crystalline contribution ($C_{IC}$).

For both $j \parallel [\overline{1}00]$ and $j \parallel [010]$ in the switched and relaxed state (\autoref{fig4}a), the observed AMR shows the presence of twofold and fourfold terms as expected from \autoref{eqAMRxx}, %
with a relatively larger strength of fourfold crystalline terms at higher fields, consistently with the AMR measurements of 50-nm thick films in Ref.~\cite{Wang2020Spin}. (For a detailed Fourier analysis of the relaxed-state and the switched-state data, see Supplementary Figs.~S4 and S5, respectively.) 
However, whereas the switched state data follow approximately the 
same dependence as the relaxed data at low fields, they depart and go out of phase at higher fields. Despite the higher-order AMR contributions being present in both curves, the difference between the switched and relaxed state reduces to a simple $A \cos(2\psi)$ form, 
as shown in \autoref{fig4}b. This uniaxial contribution produced by switching is for both $j \parallel [\overline{1}00]$ and $j \parallel [010]$ oriented along the current (and Hall bar) direction, i.e., it is a term with the same symmetry as the twofold non-crystalline AMR contribution in \autoref{eqAMRxx}.
In \autoref{fig4}c we summarize the angular dependent resistance measurements by plotting the amplitude of the difference between the switched and relaxed state as a function of the magnetic field strength.
We note that the data show a nearly linear scaling with the field, and the change of sign for $j \parallel [010]$. This change indicates the presence of several competing MR effects with a distinct field dependence.

\subsection*{C.~Pulsed-field magnetoresistance up to $\sim60$~T}
Next, we study magnetotransport of CuMnAs at extremely high magnetic fields where we exposed the samples to the magnetic field pulses up to $\sim60$~T.
The earlier work in Ref.~\cite{Wang2020Spin} reported magnetoresistances of 10-nm thick CuMnAs films at magnetic fields up to 30~T. These have shown a behaviour typical for uniaxial systems with a spin-flop transition at $\approx 2$~T and smoothly varying MR at high fields. 

Recently, magnetoresistance effects on the order of $\sim 1\%$ have been observed in antiferromagnetic Mn$_2$Au at low temperatures and at fields up to 60~T \cite{Bodnar2020Magnetoresistancea}. A transient abrupt change of MR in Mn$_2$Au observed above $\simeq 30$~T in both easy and hard field directions
was attributed to removing the domain walls by the forced reorientation of magnetic moments and the associated reduction of the domain wall contribution to the total resistance. Moreover, the persistent field-induced AMR contribution to MR on the order of $\sim 0.1\%$ was also recorded in this material. 
These findings were supported by subsequent XMLD-PEEM imaging of the magnetic structure,  
which showed a reorientation of the N\'{e}el vector to the direction perpendicular to the field, and a corresponding redistribution of magnetic domains and their sizes when exposed to high magnetic fields \cite{Sapozhnik2018Direct}.

In \autoref{fig5}, we present the high-field pulsed magnetoresistance measurements of CuMnAs in the relaxed state (without prior switching) and in the switched state. In both cases, the high-field data show a smooth variation with overlapping MRs for both the increasing and decreasing field. Also, the same zero-field resistance is measured before and after the magnetic field pulse.
This suggests that the erasing of magnetic domain walls analogous to that in Mn$_2$Au is possible neither in the relaxed nor in the switched state of CuMnAs and that the observed high-field MR evolution is related to a continuous spin reorientation within domains and
other ordinary MR effects, in agreement with Ref.~\cite{Wang2020Spin}.
Moreover, the hysteretic behaviour observed at the lower quasistationary fields (\autoref{fig3new}) is not further promoted by the $\sim60$~T field pulses
implying that the fields below 14~T are already sufficient for the saturation of the hysteresis
 (see Figs.~\ref{fig5}b,~c for comparison of the corresponding low-field data after comparable switching).
Additional temperature-dependent relaxed-state MR data at pulsed fields of various orientations (Supplementary Fig.~S6) display the largest overall MR change ($\sim8 \%$) at $\sim2$~K. 
At higher temperatures, the relative magnitude of MR 
drops down and no abrupt magnetoresistive changes are detected.

\section{IV.~Discussion} 
In this article, we studied the magnetotransport of CuMnAs in the quench-switched and relaxed state. Let us now focus on the two main findings of our measurements.
First, the quenched high-resistive state exhibits pronounced hysteretic effects and changes of slope in the magnetoresistance (\autoref{fig3new}). 
Second, the increase of the resistance of the switched state
cannot be removed by the magnetic field even when the field reaches 60~T (\autoref{fig5}). In the following paragraphs, we discuss these results in the context of the recently published microscopy \cite{krizek2020atomically} and transport measurements \cite{Kaspar2020Quenching}.

We start from the observation that the large magnetic fields do not significantly alter the increase of the resistance induced by the quench switching. Previous XMLD-PEEM and magnetoresistance measurements \cite{Wang2020Spin} showed that the magnetic field can reorient the N\'{e}el vector and drive 90$^{\circ}$ domain walls in CuMnAs. In contrast, our study of the quench switching signal shows that this mechanism is insensitive to fields that are significantly larger than the N\'{e}el vector reorientation (domain wall motion) fields.
Furthermore, recent DPC-STEM measurements have revealed additional antiferromagnetic textures in CuMnAs in the form of atomically sharp 180$^{\circ}$ domain walls \cite{krizek2020atomically}.
Combined with our high-field MR measurements, we surmise that these sharp domain walls can be responsible for the high resistance of the quench-switched state.
With the spin reversal occurring abruptly between two neighbouring magnetic sites, the atomically sharp 180$^{\circ}$ domain walls are distinct from the 90$^{\circ}$ (or 180$^{\circ}$) domain walls observed in XMLD-PEEM measurements whose width is approximately 100~nm.
Since the two domains separated by the atomically sharp 180$^{\circ}$ domain wall have antiparallel N\'{e}el vectors, they are rotated coherently by the magnetic field, leaving the sharp 180$^{\circ}$ domain wall unaffected. In other words, the atomically sharp 180$^{\circ}$ domain walls cannot be manipulated by the spin reorientation mechanism \cite{Gomonay2016High}. On the other hand, they can readily explain the large and field-insensitive quench-switching resistive signal.

Within the picture of the coexisting wide and atomically sharp domain walls, we can also explain the observed hysteresis of MR in the quench-switched state. 
We consider that the intertwined atomically sharp domain walls influence
the N\'{e}el vector reorientation and the motion of the wide domain walls.
The atomically sharp domain walls at high enough densities to generate the sizeable resistive switching signals can serve as pinning centres for the wide 90$^{\circ}$ 
domain walls and hinder their motion. Correspondingly, the magnitude of the hysteretic MR reaches $\sim 0.1$ \%, consistent with the observed AMR scale in CuMnAs \cite{Wang2020Spin}. 
At lower densities of the sharp domain walls (and corresponding lower quench-switching signals $\lesssim 20\%$), this effect is less pronounced and the MR is only slightly modified compared to the relaxed state.

\section{V.~Summary}

In summary, we report hysteretic phenomena and slope changes in the magnetoresistance of CuMnAs when the sample is quench-switched into high-resistive states by electrical pulses. The magnitude of the effects scales up with the increasing resistive quench-switching signal. We also observe that, apart from the weak and transient MR effects, the high-resistive quench-switched state cannot be erased by a magnetic field as high as 60~T.
We interpret the observed MR phenomena in terms of an interplay 
of wide 90$^{\circ}$ domain walls and atomically sharp 180$^{\circ}$ domain walls,  reported earlier in  XMLD-PEEM and DPC-STEM imaging experiments in CuMnAs.

\section{Acknowledgements}
This work was supported by the Ministry of Education of the Czech Republic Grants  LNSM-LNSpin, LM2018140, Czech Science Foundation Grants No.  21-28876J, and the EU FET Open RIA Grant No. 766566.
Experiments at magnetic fields up to 14~T were performed in MGML (mgml.eu), which is supported within the program of Czech Research Infrastructures (project no. LM2018096). For the pulsed-field measurements, we acknowledge the support of the HLD at HZDR, member of the European Magnetic Field Laboratory (EMFL). We also acknowledge fruitful discussions with K.~V\'yborn\'y during the preparation of the manuscript. 

\bibliography{mybibliography_spintronics}

\onecolumngrid


\newpage

\newpage
\color{white}
\subsection*{Fig. 2}
\color{black}

\newpage
\color{white}
\subsection*{Fig. 3}
\color{black}

\newpage
\color{white}
\subsection*{Fig. 4}
\color{black}
\begin{figure}[h]
 \centering
 \includegraphics[width=\textwidth]{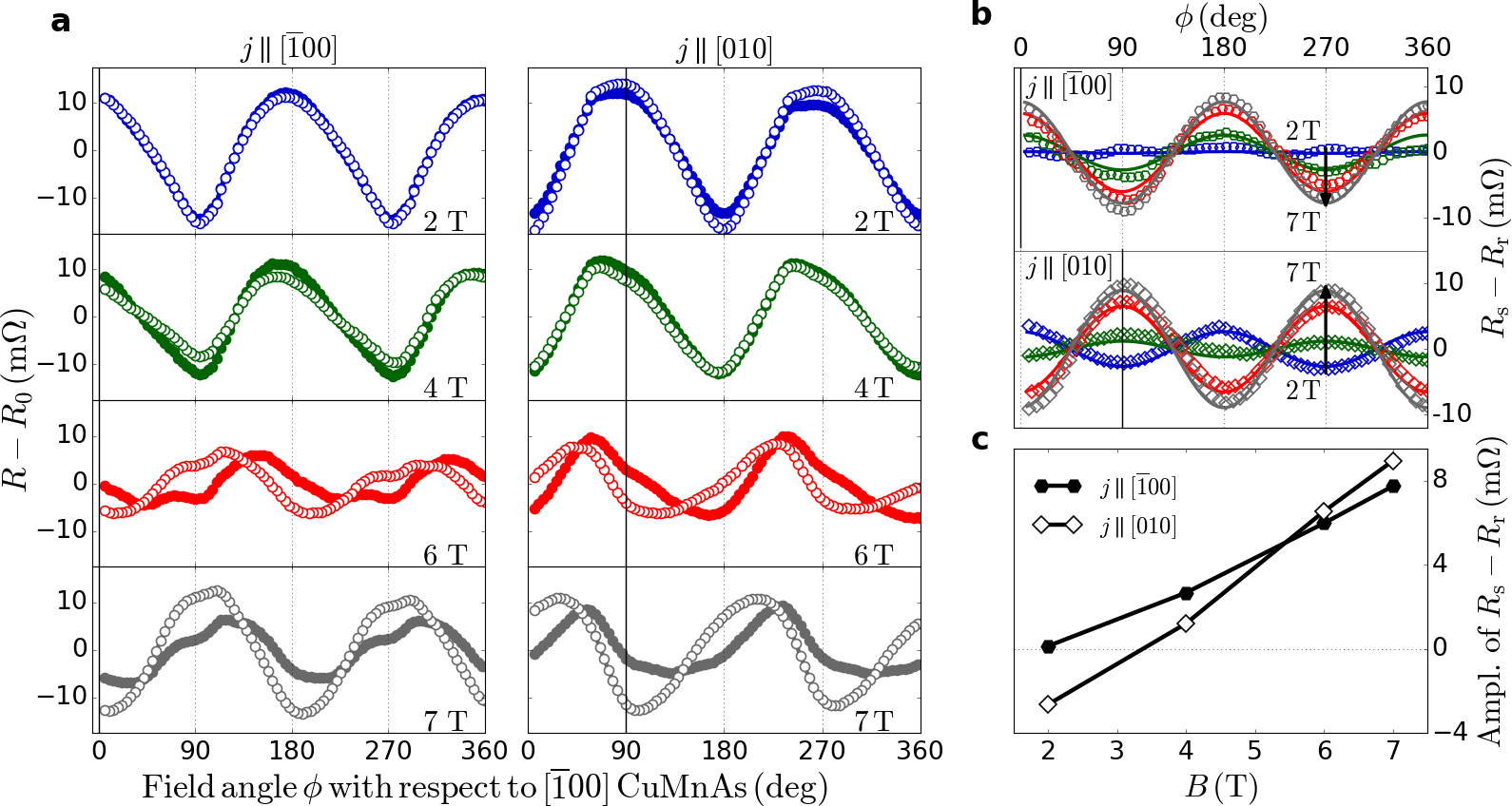}
 \caption{\textbf{(a)}~Resistance at 2~K during magnetic field rotations at various field strengths for $j\parallel[\overline{1}00]$ and $j\parallel[010]$ current directions. Full symbols represent the MR data in the switched state after the 30~V electrical pulse at 200~K, open symbols stand for the MR in the relaxed state.  Full vertical lines indicate the current direction. The individual angular scans were collected at increasing fields from 2 to 7~T; the data for the field rotation from 0 to 360$^{\circ}$ coincided with that taken during rotation from 360$^{\circ}$ back to 0 (not shown). \textbf{(b)}~Difference between the switched and relaxed MR data for increasing field strengths. The full lines represent fits to the $A\cos(2\psi)$ function. \textbf{(c)}~The amplitude of differential resistance as obtained by fitting to the $A\cos(2\psi)$ function in~\textbf{(b)} above.   
 }
 \label{fig4}
\end{figure}
\newpage
\color{white}
\subsection*{Fig. 5}
\color{black}
\begin{figure}[h]
 \centering
\includegraphics[width=0.5\textwidth]{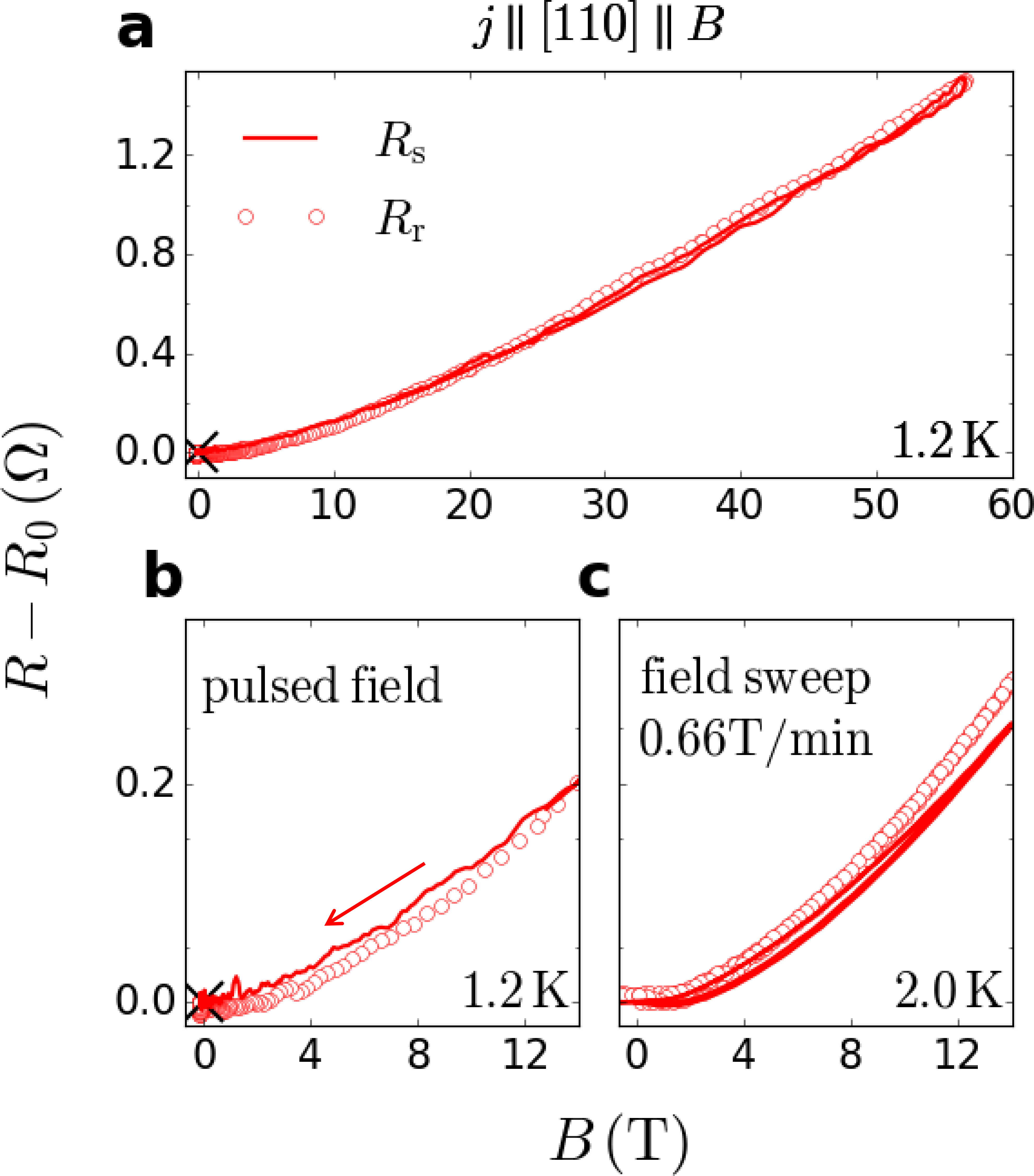} 
 \caption{\textbf{(a)}~Magnetoresistance of CuMnAs in the pulsed magnetic field at low temperature. The switched-state MR ($R_{\mathrm{s}}$) for the switching signal 7.2~$\Omega$ is plotted with a full line, open symbols stand for the MR in the relaxed state ($R_{\mathrm{r}}$).
 Data from the sweep up below 10~T had to be omitted due to the large noisy pickup signal (typical for the given type of measurement). The resistance $R_{\mathrm{0}}$ in the switched state at zero field immediately before the magnetic field pulse is highlighted with the black cross symbol ($\times$) . 
 \textbf{(b)}~Detail of the pulsed-field MR from \textbf{(a)} below 14~T.
 \textbf{(c)} MR data obtained at a field sweep rate of 0.66~T/min without switching (open symbols) and after electrical switching with the signal magnitude of 6.8~$\Omega$ (full line).
 }
 \label{fig5}
\end{figure}


\end{document}


\title{Supplemental Material: Hysteretic effects and magnetotransport of electrically switched CuMnAs}

\author{Jan~Zub{\'a}{\v c}}
\affiliation{Institute of Physics, Czech Academy of Sciences,
Cukrovarnick{\'a} 10, 162 00, Prague 6, Czech Republic}
\affiliation{Charles University, Faculty of Mathematics and Physics, Ke Karlovu 3, 121 16 Prague 2, Czech Republic}

\author{Zden{\v e}k~Ka{\v s}par}
\affiliation{Institute of Physics, Czech Academy of Sciences,
Cukrovarnick{\'a} 10, 162 00, Prague 6, Czech Republic}
\affiliation{Charles University, Faculty of Mathematics and Physics, Ke Karlovu 3, 121 16 Prague 2, Czech Republic}

\author{Filip~Krizek}
\affiliation{Institute of Physics, Czech Academy of Sciences,
Cukrovarnick{\'a} 10, 162 00, Prague 6, Czech Republic}

\author{Tobias~F{\"o}rster}
\affiliation{Hochfeld-Magnetlabor Dresden (HLD-EMFL) and W\"urzburg-Dresden Cluster of Excellence ct.qmat, Helmholtz-Zentrum
Dresden-Rossendorf, 01328 Dresden, Germany}

\author{Richard~P.~Campion}
\affiliation{School of Physics and Astronomy, University of Nottingham,Nottingham NG7 2RD, United Kingdom}

\author{V{\'i}t~Nov{\'a}k}
\affiliation{Institute of Physics, Czech Academy of Sciences,
Cukrovarnick{\'a} 10, 162 00, Prague 6, Czech Republic}

\author{Tom{\'a}{\v s}~Jungwirth}
\affiliation{Institute of Physics, Czech Academy of Sciences,
Cukrovarnick{\'a} 10, 162 00, Prague 6, Czech Republic}
\affiliation{School of Physics and Astronomy, University of Nottingham,Nottingham NG7 2RD, United Kingdom}

\author{Kamil~Olejn{\'i}k}
\affiliation{Institute of Physics, Czech Academy of Sciences,
Cukrovarnick{\'a} 10, 162 00, Prague 6, Czech Republic}

\date{\today}
%
%
%
%
%
\maketitle
%
%

\beginsupplement
\newcounter{suppfig}
\renewcommand{\thefigure}{S\arabic{figure}}%

\begin{figure}[hb]
 \centering
 \includegraphics[width=0.5\textheight]{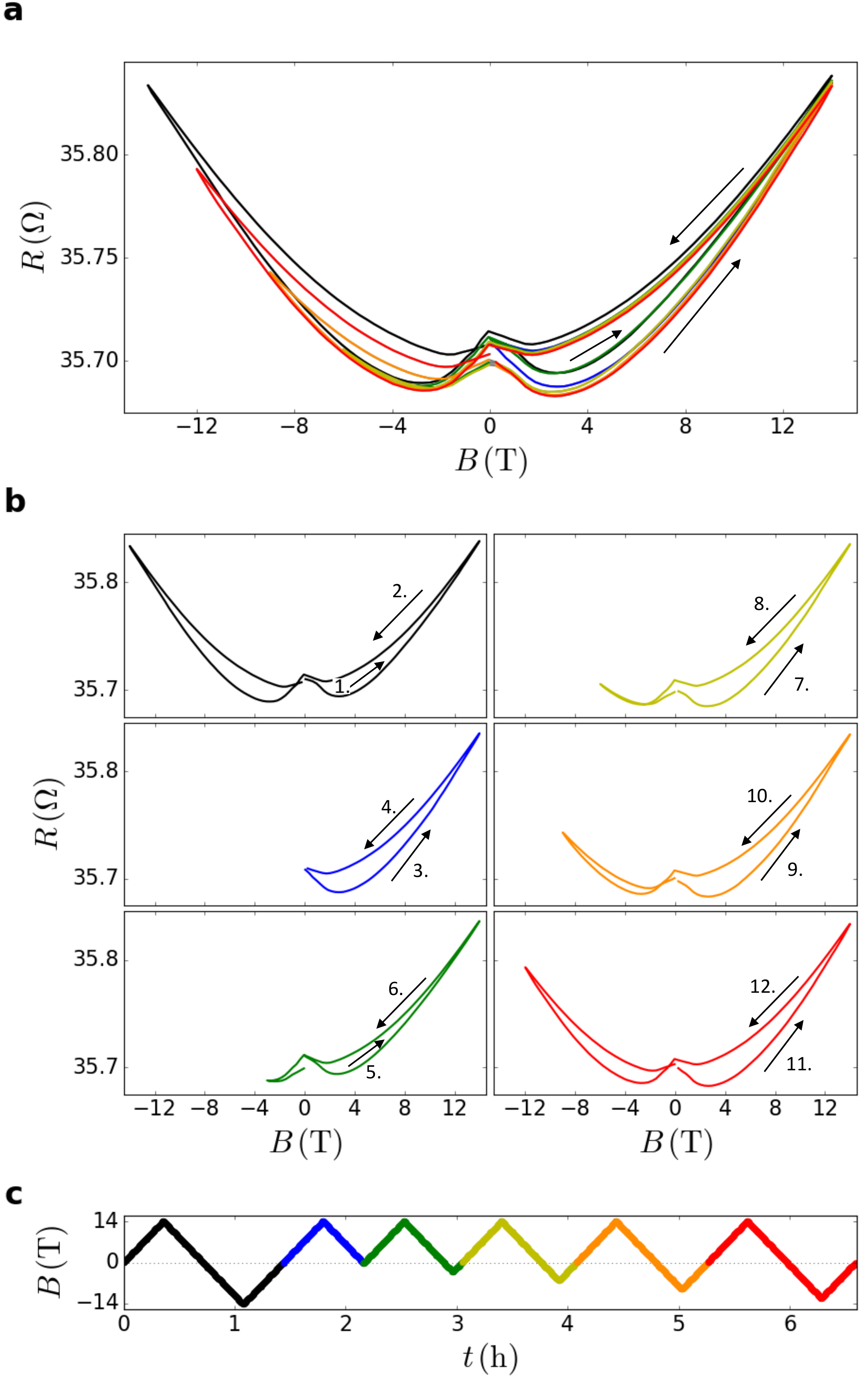}
 \caption{Details of hysteretic behaviour of CuMnAs in the switched state.  \textbf{(a)} Consecutive MR loops in the switched state for the maximum field of 14 T and different minimal negative fields.    \textbf{(b)} Same data as in \textbf{(a)} but decomposed into individual loops. \textbf{(c)} Temporal dependence of the magnetic field for the data in \textbf{(a, b)} (colours correspond to those used above). 
 }
\label{figS1}
\refstepcounter{suppfig}\label{suppfigS1}
\end{figure}

\newpage
\begin{figure}[h]
 \centering
 \includegraphics[width=\textwidth]{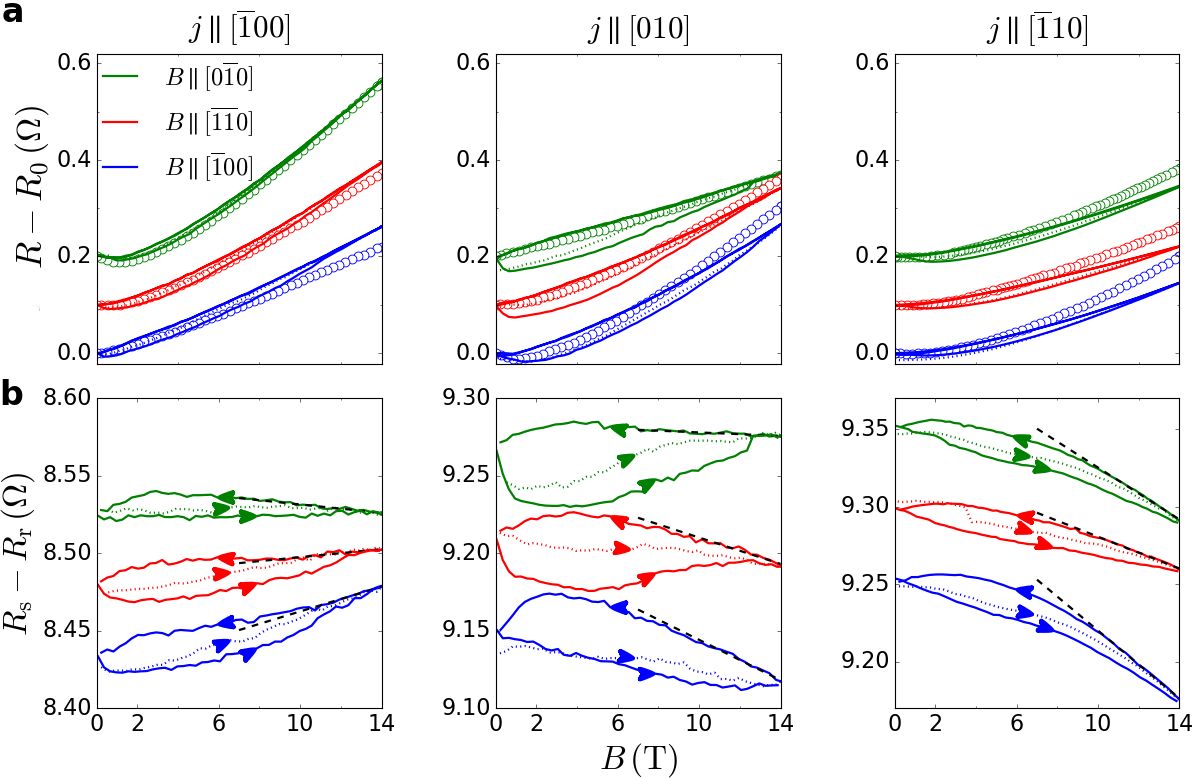}
 \caption{ 
 		\textbf{(a)}~Magnetoresistance of CuMnAs at 2~K in the switched state for the signal
 		$R_{\mathrm{s}}~-~R_{\mathrm{r}}\approx 9\, \Omega$ (full line) and in the relaxed state (empty symbols) for various mutual current and field configurations. Pulsing and probing current directions were identical.
 		\textbf{(b)}~Difference between switched ($R_{\mathrm{s}}$) and relaxed ($R_{\mathrm{r}}$) magnetoresistance
 		showing hysteretic behaviour. The curves were obtained by subtracting corresponding MR dependencies in \textbf{(a)}.
 		The intercepts of y-axes illustrate the absolute magnitude of the switching-induced signal.  
 		The dashed black lines show the high-field slope of the difference curves obtained from the sweep down. 
 		The green and red curves are offset by an arbitrary value for lucidity.}
    \label{figS2}
    \refstepcounter{suppfig}\label{suppfigS2}
\end{figure}

\newpage
\begin{figure}[h]
 \centering
\includegraphics[width=\textwidth]{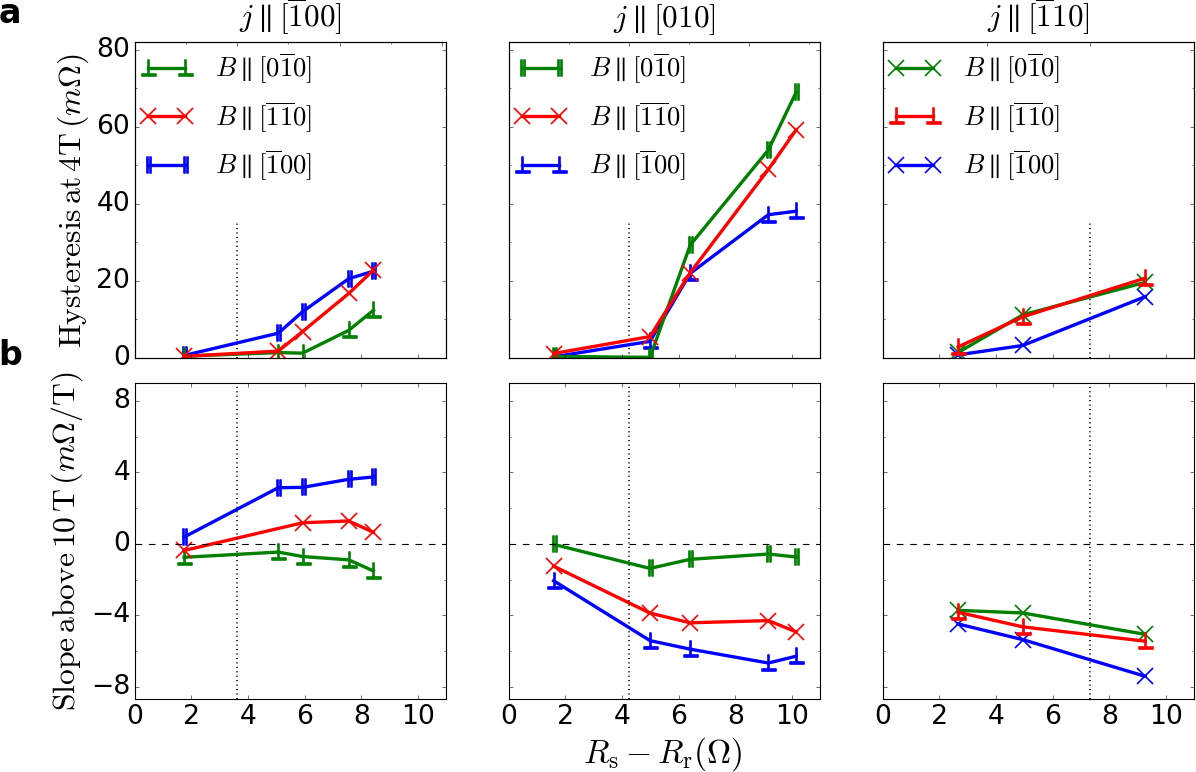}
 \caption{Parameters of the difference MR curves as a function of the switching signal for various mutual current and field configurations:
 \textbf{(a)}~ Hysteresis magnitude evaluated at 4~T. \textbf{(b)}~Slope of the difference curve (sweep down) as obtained from a linear fit in the range 10-14~T. Symbols $\parallel$, $\perp$ and $\times$ were used to signify whether the magnetic field was applied (anti)parallel, perpendicular or at the angle of 45$^{\circ}$(135$^{\circ}$) with respect to the readout electrical current. The dotted vertical lines represent the relative switching signal $\frac{  R_{\mathrm{s}} - R_{\mathrm{r}}   }{  R_{\mathrm{2K}}  } = 20\%$. }
		
    \label{figS3}
    \refstepcounter{suppfig}\label{suppfigS3}
\end{figure}


\begin{figure}[h!]
 \centering
 \includegraphics[height=0.8\textheight]{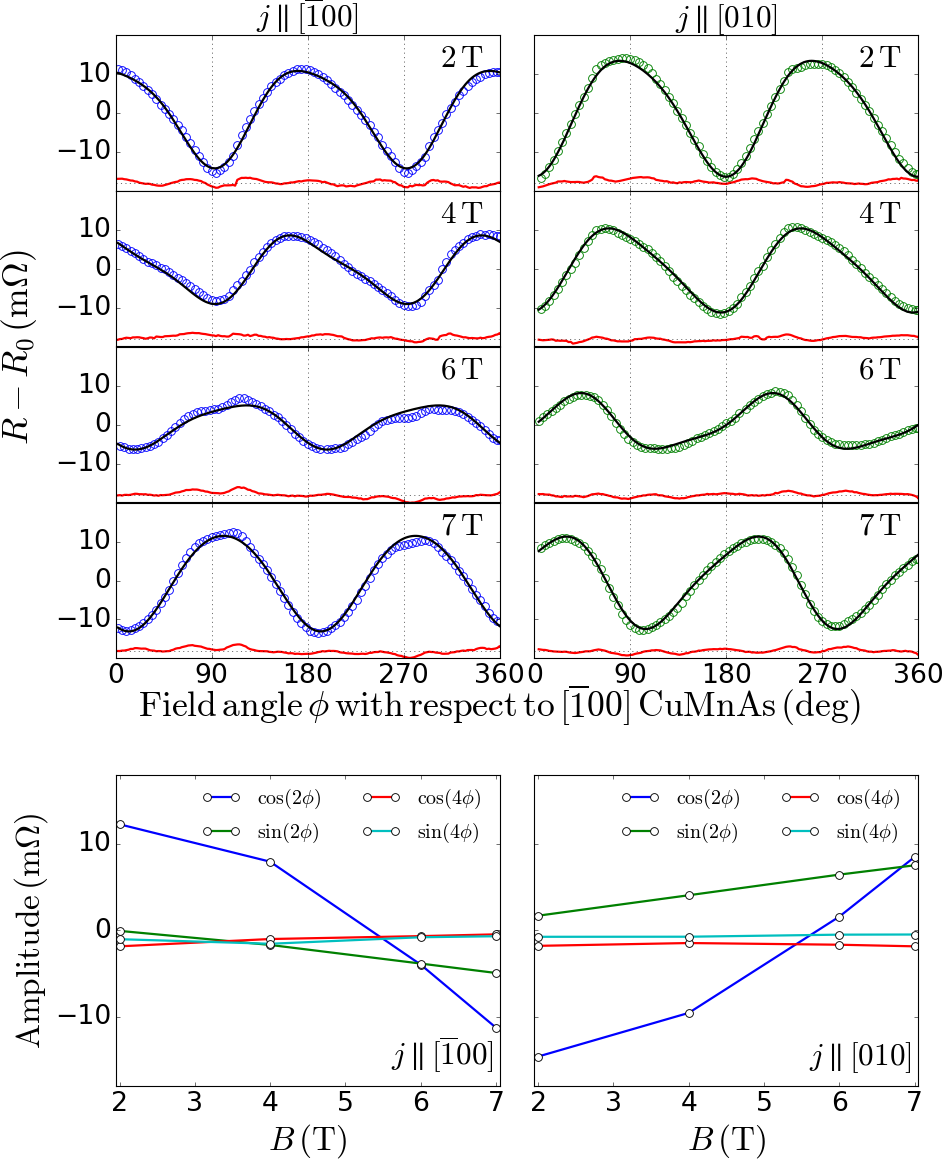}
  \caption{Analysis of the angular resistance scans in the relaxed state. In the top panel we show the experimental data (open symbols) for the  $j \parallel [\overline{1}00]$ (blue) and  $j \parallel [010]$ (green)  and varying field strengths. The black line represents a fit of the data to the function $f(\phi) = A_1\cos(2\phi) + A_2\sin(2\phi) + A_3\cos(4\phi) + A_4\sin(4\phi) $, the red line is the shifted difference between the data and the fit. In the bottom panel, we plot the obtained amplitudes as a function of the magnetic field.}
\label{figS4_1}
\refstepcounter{suppfig}\label{suppfigS4}
\end{figure}

\begin{figure}[h!]
 \centering
 \includegraphics[height=0.8\textheight]{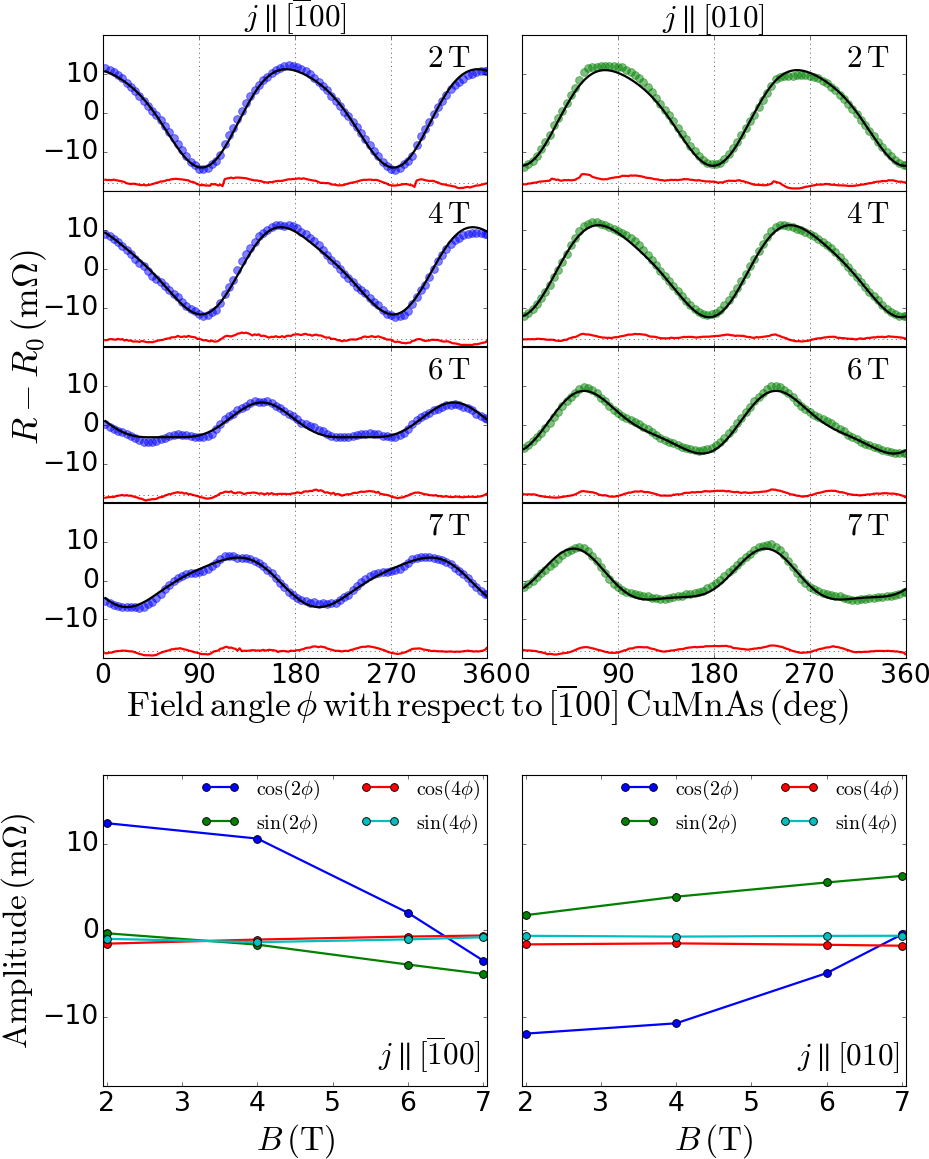}
   \caption{Analysis of the angular resistance scans in the switched state. In the top panel we show the experimental data (full symbols) for the  $j \parallel [\overline{1}00]$ (blue) and  $j \parallel [010]$ (green)  and varying field strengths. Black line represents a fit of the data to the function $f(\phi) = A_1\cos(2\phi) + A_2\sin(2\phi) + A_3\cos(4\phi) + A_4\sin(4\phi) $, the red line is the shifted difference between the data and the fit. In the bottom panel, we plot the obtained amplitudes as a function of magnetic field.}
\refstepcounter{suppfig}\label{suppfigS5}
\end{figure}

\newpage
\begin{figure}[h]
 \centering
 \includegraphics[width=\textwidth]{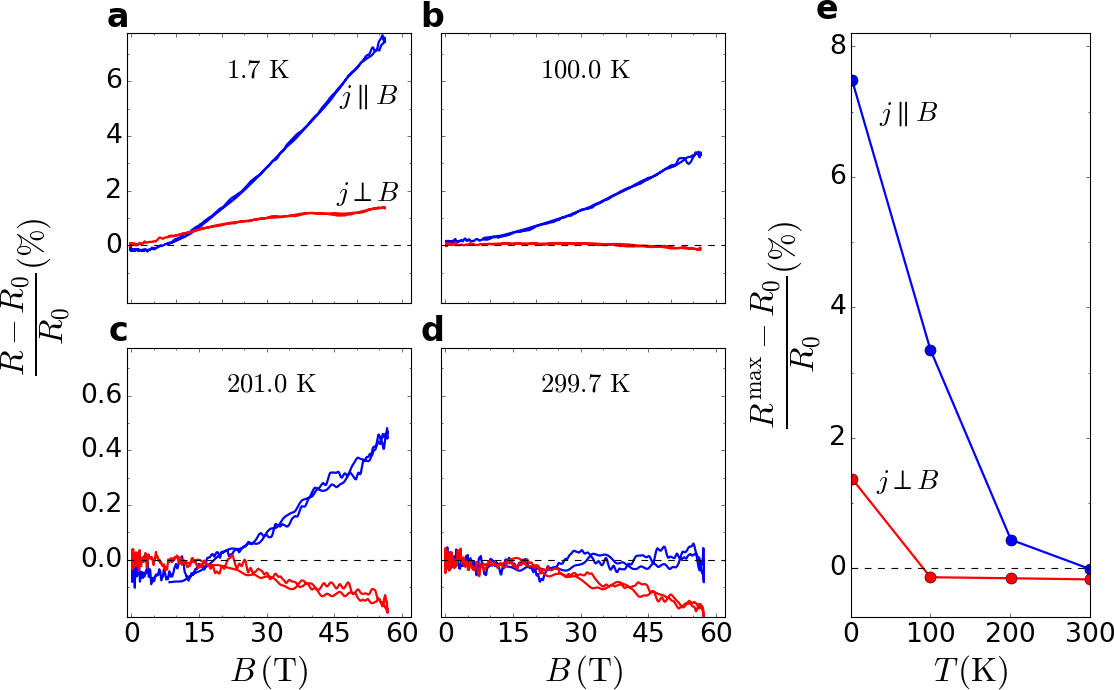}
 \caption{Temperature-dependent magnetoresistance of pristine CuMnAs for $j \parallel [\overline{1}10]$ under pulsed magnetic field. 
 \textbf{(a-d)} The MR data during the magnetic field pulse applied parallel (blue) and perpendicular (red) to the current direction. Both up and down field sweeps are presented. The data from the sweep up below $\approx 10$~T are omitted due to the additional random pickup signal, which is inevitable in the given experimental arrangement. Please note that the y-axes in \textbf{(a, b)}
 are scaled by a factor of 10 with respect to those of  \textbf{(c, d)} due to relatively higher MR at the lower temperatures. 
 \textbf{(e)} The MR at the maximum field of 56~T as a function of temperature. 
 }
 \refstepcounter{suppfig}\label{suppfigS6}
\end{figure}

\renewcommand{\thefigure}{S\arabic{figure}}%
\begin{figure}[h]
 \centering
 \includegraphics[width=0.9\textwidth]{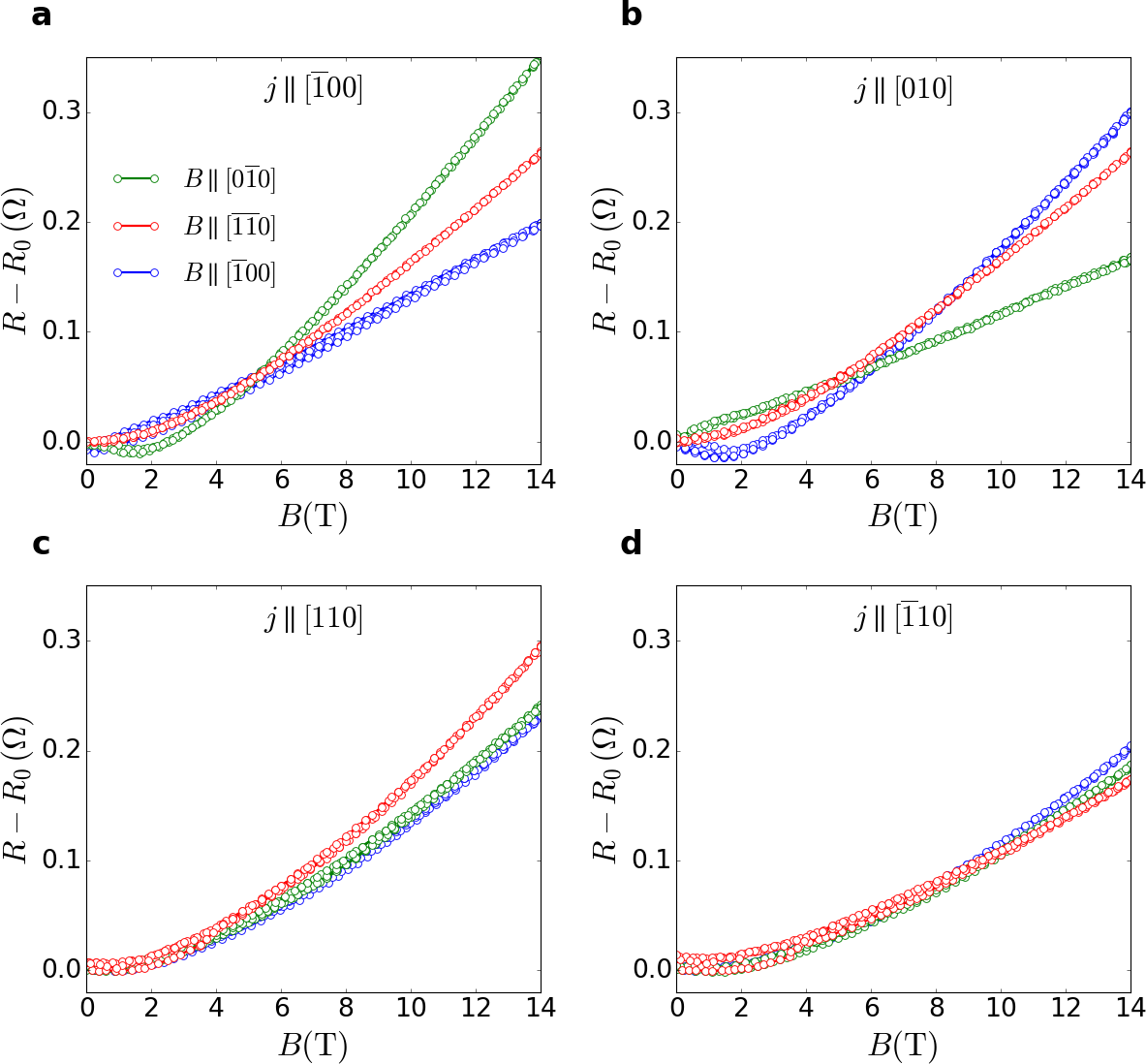}
 \caption{Magnetoresistance of 50~nm CuMnAs epilayer measured in the relaxed state at 2~K. The feature characteristic for continuous spin reorientation in biaxial CuMnAs films can be observed at 2~T \textbf{(a)} for $j \parallel [\overline{1}00]$ with $B \parallel [0\overline{1}0]$ and \textbf{(b)} for
 $j \parallel [010]$ with $B \parallel [\overline{1}00]$ and $B \parallel [0\overline{1}0]$, respectively. Data from several field sweeps are presented.}
\label{figS4}
\end{figure}
